\documentclass[aps,prl,reprint,superscriptaddress]{revtex4-2}

\usepackage[utf8]{inputenc}
\usepackage{amsthm}
\usepackage{amsmath}
\usepackage{amsfonts}
\usepackage{color}
\usepackage{dsfont}
\usepackage{dsfont}
\usepackage{soul}
\usepackage{hyperref}
\usepackage{graphicx,xcolor}
\usepackage{braket}
\usepackage{comment}

\newcommand{\bcdot}{\boldsymbol{\cdot}}
\newcommand{\dket}[1]{|#1\rangle\!\rangle} 

\begin{document}

\newcommand{\liege}{Institut de Physique Nucléaire, Atomique et de Spectroscopie, CESAM, Universit\'e de Li\`ege, 4000 Liège, Belgium}

\title{Controlling Matter Phases beyond Markov}

\author{Baptiste Debecker}
\affiliation{\liege}
\author{John Martin}
\affiliation{\liege}
\author{François Damanet}
\affiliation{\liege}

\begin{abstract}
Controlling phase transitions in quantum systems via coupling to reservoirs has been mostly studied for idealized memory-less environments under the so-called Markov approximation. Yet, most quantum materials and experiments in the solid state, atomic, molecular and optical physics are coupled to reservoirs with finite memory times. Here, using the spectral theory of non-Markovian dissipative phase transitions developed in the companion paper~[Debecker~\textit{et al.}, Phys.\ Rev.\ A \textbf{110}, 042201 (2024)], we show that memory effects can be leveraged to reshape matter phase boundaries, but also reveal the existence of dissipative phase transitions genuinely triggered by non-Markovian effects.
\end{abstract}

\date{\today}
\maketitle

\emph{Introduction.} Finding new ways to control phase transitions in quantum systems to access desired properties is at the forefront of research for developing new materials and technologies. In this context, driven-dissipative mechanisms obtained via the coupling of systems to engineered environments and fields offer opportunities to generate matter phases otherwise inaccessible~\cite{Toner1998Flocks, Jin2016Cluster, Lee2013Unconventional}.

However, thus far, dissipative phase transitions (DPTs), which have been observed in controlled experiments~\cite{Baumann2010Apr, Klinder2015Dynamical, Fitzpatrick2017Observation, Fink2017Observation, Fink2017Signatures, Benary2022, Beaulieu2023Observation, Ferri2021Emerging, Berdou2023One}, have mostly garnered theoretically attention in systems coupled to memoryless reservoirs~\cite{Minganti2018Spectral, Kessler2012Dissipative, Hwang2018Dissipative}. 
Yet, most realistic systems are coupled to reservoirs with a spectral structure~\cite{deVega2017Dynamics}, giving the latter a memory of past system-bath exchanges, which considerably complicates their dynamics. Such non-Markovian effects are crucial to be understood, not least because they can be used as a resource to generate useful phenomena, such as non-Markovian-assisted steady state entanglement~\cite{Huelga2012Non}, quantum transport~\cite{Maier2019Environment}, spin squeezing~\cite{Link2022}, chaotic behaviors~\cite{Chen2023C} or new dynamical phases~\cite{Otterpohl2022Hidden}. Moreover, from a computational perspective, it is sometimes desirable to derive reduced descriptions of a large Markovian open quantum system in order to deal with a smaller Hilbert space, which usually implies dealing with non-Markovian effects~\cite{Damanet2019, Palacino2020}.

Here, using the theoretical framework developed in the companion paper~\cite{Debecker2024PRA} which generalizes the spectral theory of dissipative phase transitions to non-Markovian systems, we highlight how non-Markovian effects can be used to reshape phase boundaries but also capture DPTs totally overlooked by the Markov approximation, which opens many possibilities to explore DPTs in a wider range of systems and to access matter phases in new regimes of parameters and experimental setups. By contrast with previous studies of non-Markovian effects in DPTs via other techniques~\cite{Nagy2015Nonequilibrium,Nagy2016Critical, Haikka2012NonMarkovianity,Strathearn2018Efficient}, which mostly focus on the paradigmatic spin-boson model~\cite{Anders2007, Chin2011Generalized, Florens2010Quantum}, here we provide a general method, which we apply in two specific examples.

Below, we first summarize the theoretical framework presented in~\cite{Debecker2024PRA} and in particular its central element: the generalization of the Liouvillian for non-Markovian systems, its properties and its connections with DPTs and symmetries. Then, we apply the method to a generalized Lipkin-Meshkov-Glick model~\cite{Lee2014D} and show that deviations from a Markovian reservoir lead to a shift of the phase transition boundary. Then, even more remarkably, we reveal the existence of DPTs triggered by non-Markovianity.

\begin{figure}
    \centering
     \includegraphics[width=0.475\textwidth]{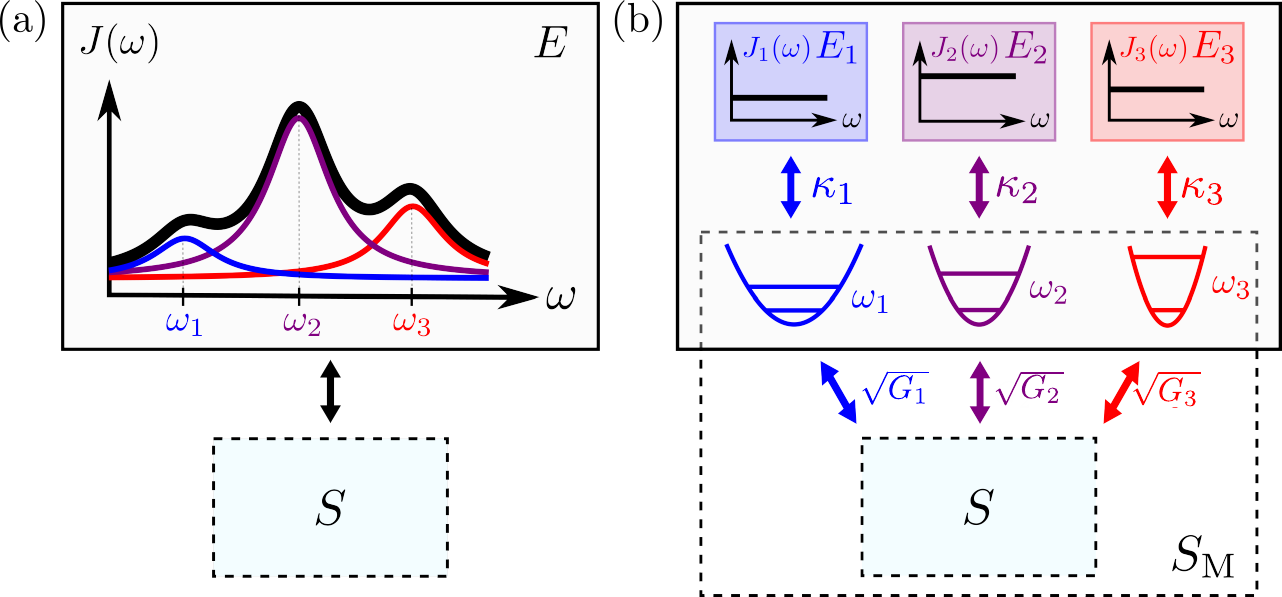}
    \caption{(a): Sketch of a system $S$ interacting with a structured environment $E$ characterized by a spectral density $J(\omega)$ that can be decomposed into three Lorentzians, as if the system was coupled to three pseudo-modes coupled to their own unstructured bath (b). If one enlarges the system $S$ by including the pseudo-modes (system $S_M$ in the dashed black box), the dynamics can be treated by a Lindblad description.}
    \label{Pseudomodes}
\end{figure}

\emph{Theoretical framework.} 
Consider a system $S$ coupled to a bosonic environment $E$ at zero temperature~\cite{footnote_formalism}. The total Hamiltonian ($\hbar = 1)$ is
\begin{equation}
    H = H_S + \underbrace{\sum_k \omega_k a_k^\dagger a_k}_{\equiv~H_E} + \underbrace{\sum_{k} (g_{k}a_k L^\dagger + g_{k}^* a_k^\dagger L)}_{\equiv~H_{\mathrm{int}} },
\label{GeneralHamiltonian}
\end{equation}
where $H_S$ ($H_E$) is the system (environment) Hamiltonian, with $a_k$ ($a_k^\dagger$) the annihilation (creation) operator for the $k$-th mode of frequency $\omega_k$, and $H_\mathrm{int}$
 is the interaction Hamiltonian with $L$ being an arbitrary system operators and $g_k$ being the system-bath coupling strengths. The effect of the environment on the system is encoded in the spectral density $J(\omega) = \pi \sum_k |g_k|^2 \delta(\omega-\omega_k)$ or equivalently in the bath correlation function $\alpha(\tau) = \sum_k |g_k|^2  e^{-i \omega_k \tau} = (1/\pi)\int_0^\infty J(\omega) e^{-i\omega \tau} \mathrm{d}\omega$. We assume that the correlation function, which depends on the model, is a sum of $M$ decaying exponentials
\begin{equation}
    \alpha(\tau) =  \sum_{j=1}^M G_j\, e^{-i\omega_j \tau - \kappa_j |\tau| },\quad \kappa_j, \omega_j \in \mathbb{R},\; G_j \in \mathbb{C}.
    \label{CF}
\end{equation}
This decomposition, which is not unique~\cite{Tang2015Extended}, can be performed either exactly or with great precision in a wide range of situations~\cite{Meier1999Non,Ritschel2014, Hartmann2019, Lambert2020Qutip}. For $G_j \in \mathbb{R}$, this amounts to decompose the non-Markovian environment $E$ into a set of $M$ modes of frequencies $\{\omega_j\}$ which are damped with rates $\{\kappa_j\}$ due to their coupling to independent Markovian baths, as illustrated in Fig.~\ref{Pseudomodes}. 
This pseudo-mode picture~\cite{Imamoglu1994Stochastic, Dalton2001Theory, Garraway1997Nonperturbative, Pleasance2020Generalized, Mazzola2009Pseudomodes, Yang2012Nonadiabatic,Breuer2004Genuine, Barchielli2010Stochastic} is relevant for atoms in cavities, superconducting qubits coupled to resonators~\cite{Schmidt2013CircuitQED, Blais2021Circuit}, electrons-phonon systems~\cite{Flannigan2022Manybody, Moroder2022Strongly}, or emitters in plasmonic cavities~\cite{Santhosh2015Vacuum}. Using complex $G_j$ is even more general and allows for instance an efficient fit of Ohmic spectral density~\cite{Mascherpa2020, Tang2015Extended} and the study of critical behaviors \cite{Duan2017Zero}.

When the global system is initially in the state $\rho(0) = \rho_S(0) \otimes \rho_B(0)$, the exact dynamics of $S$ can be described by the HEOM method which takes the form~\cite{Tanimura89, Ishizaki2005, Ishizaki2009, Yoshitaka2020, Lambert2020Qutip}
\begin{align}
    \frac{d\rho^{(\vec{n}, \vec{m})}}{dt} &= -i[H_S, \rho^{(\vec{n}, \vec{m})}] - (\vec{w}^*\bcdot\vec{n} + \vec{w}\bcdot\vec{m}) \rho^{(\vec{n}, \vec{m})} \nonumber\\
    &+ \sum_{j = 1}^M \left\{ G_j n_j L \rho^{(\vec{n}-\vec{e_j}, \vec{m})} + G_j^* m_j \rho^{(\vec{n}, \vec{m}-\vec{e_j})}L^\dagger \right. \nonumber\\
    &\qquad \quad  + \left.[\rho^{(\vec{n}+\vec{e_j}, \vec{m})}, L^\dagger] + [L, \rho^{(\vec{n}, \vec{m}+\vec{e_j})}]\right\},
 \label{HEOM}
\end{align}
with $\vec{n} = (n_j)$ and $\vec{m}= (m_j)$ multi-indices in $\mathbb{N}^M$, $\vec{w} = (\kappa_j + i\, \omega_j) \in \mathbb{C}^M$, $\vec{e_j} = (\delta_{jj'})$ unit vectors, and $\vec{a}\bcdot \vec{b} = \sum_j a_j^* b_j$ the inner product on $\mathbb{C}^M$.
In Eq.~(\ref{HEOM}), $\rho^{(\vec{0}, \vec{0})} \equiv \rho_S$ corresponds to the physical density operator of the system $S$ with which all the mean values of system observables are computed, while $\rho^{(\vec{n}, \vec{m})}$ for $(\vec{n}, \vec{m}) \neq (\vec{0},\vec{0})$, which are also operators acting on the system space, correspond to auxiliary states from which bath correlations can be obtained~\cite{Link2022}. 
Although the hierarchy is formally infinite, it can be truncated at large hierarchy depth indices $\vec{n}$ and $\vec{m}$.
In practice, the stronger the non-Markovianity, the larger the number of auxiliary states we need to retain to obtain convergence of the results. Here, we choose the triangular truncation $\rho^{(\vec{n}, \vec{m})} = 0 \; \forall \:\vec{n},\vec{m}: \sum_j (n_j + m_j) > k_\mathrm{max}$, where $k_\mathrm{max}$ is the truncation order,
yielding a total of $K = (2M+k_{\mathrm{max}})!/((2M)!\, k_{\mathrm{max}}!)$ auxiliary states~\cite{Link2022}.

By stacking in a vector $\dket{\rho}$ all the vectorized versions of the matrices $\rho^{(\vec{n}, \vec{m})}$, Eq.~\eqref{HEOM} can be written in the form
\begin{equation}
    \frac{d\dket{\rho}}{dt} = \mathcal{L}_{\mathrm{HEOM}}(k_{\mathrm{max}})\,\dket{\rho},
    \label{def_Leff}
\end{equation}
where $\mathcal{L}_{\mathrm{HEOM}}(k_{\mathrm{max}})$ is the \textit{HEOM Liouvillian}, the generator of the non-Markovian dynamics of the system, exact for $k_\mathrm{max} \to +\infty$ and which generalises Lindblad's Liouvillian. Instead of using $\mathcal{L}_{\mathrm{HEOM}}$, one can sometimes as noted above enlarge the system by including explicit pseudo-mode degrees of freedom damped by standard Lindblad decay channels, as illustrated in Fig.~\ref{Pseudomodes}(b). This would define a Markovian Liouvillian $\mathcal{L}_{M}$ for the global system $S_M$. However, as shown in~\cite{Debecker2024PRA} using $\mathcal{L}_{\mathrm{HEOM}}$ is computationally more favorable than $\mathcal{L}_{M}$, especially for large $M$.

\emph{Properties of the HEOM Liouvillian.} The superoperator
$\mathcal{L}_\mathrm{HEOM}$ is linear and in general non-Hermitian. We assume it is diagonalizable and denote its eigenvectors and eigenvalues by $\dket{\rho_i}$ and $\lambda_i$. For a truncation order $k_{\mathrm{max}}$, its dimension is $
D = K\,\mathrm{dim(\mathcal{H}_S})^2$. It admits the following properties: (i) its spectrum is symmetric with respect to the real axis; (ii) it preserves the trace of the physical state $\rho^{(\vec{0},\vec{0})}$; (iii) the eigenvalue $0$ is always in its spectrum, guaranteeing the existence of a stationary state; (iv) all the eigenvalues must have a negative real part in the limit $k_{\mathrm{max}} \rightarrow +\infty$; (v) $\mathrm{Tr}[\mathds{1}^{(\vec{0}, \vec{0})}\rho_i] = 0$ with $\mathds{1}^{(\vec{0}, \vec{0})}$ the projector onto the physical state space if $\rho_i$ is a right eigenoperator of $\mathcal{L}_{\mathrm{HEOM}}$ associated with the eigenvalue $\lambda_i$ with $\mathrm{Re}[\lambda_i] \neq 0$. As in~\cite{Minganti2018Spectral, Kessler2012Dissipative}, we order the eigenvalues of $\mathcal{L}_{\mathrm{HEOM}}$ so that
$|\mathrm{Re}[\lambda_0]| < |\mathrm{Re}[\lambda_1]| < \dots < |\mathrm{Re}[\lambda_D]|$, where $\lambda_0=0$.

\emph{DPT and HEOM Liouvillian spectrum.}
Consider a system described by Eq.~\eqref{def_Leff} which admits a valid thermodynamic limit $N \to \infty$ and a unique steady state $\rho_{ss}$ for all finite $N$. We say that the system undergoes a phase transition of order $p$ when a non-analytical change in a $g$-independent system observable $O$ occurs when the parameter $g$ tends to a critical value $g_c$ for $N \to \infty$, i.e.,~\cite{Minganti2018Spectral}
 \begin{equation}\label{defDPT}
     \lim_{g\to g_c} \left|\frac{\partial^p}{\partial g^p} \lim_{N\to +\infty} \langle O \rangle_{ss}\right| = +\infty,
 \end{equation}
where $\langle O \rangle_{ss} = \mathrm{Tr}[O\rho_{ss}^{(\vec{0},\vec{0})}]$. This definition of DPTs is the same as for Markovian systems. The only difference is that the steady state is now obtained from the HEOM Liouvillian~(\ref{def_Leff}). 
Like for the Markovian case, a non-analytical change as described by~(\ref{defDPT}) must occur due to a level crossing in the spectrum of $\mathcal{L}_{\mathrm{HEOM}}$, which implies the closing of the HEOM Liouvillian gap $\mathrm{Re}[\lambda_1]$. For $1^{\mathrm{st}}$-order DPTs, the connection is even stronger as a DPT occurs \textit{iff} $\mathrm{Re}[\lambda_1] = 0$ at $g=g_c$ and $\mathrm{Im[\lambda_1]} = 0$ in a finite domain around $g_c$ for $N \to \infty$~\cite{Debecker2024PRA}.

\emph{Symmetries and DPTs.} We call \textit{weak symmetry} of $\mathcal{L}_{\mathrm{HEOM}}$ any unitary superoperator $\mathcal{U}$ such that $[\mathcal{L}_{\mathrm{HEOM}}, \mathcal{U}] =0$. The matrix representing $\mathcal{L}_{\mathrm{HEOM}}$ in the eigenvector basis of $\mathcal{U}$ is block-diagonal, i.e., $\mathcal{L}_{\mathrm{HEOM}} = \bigoplus_ {u_k} \mathcal{L}_{u_k}$, where each block $\mathcal{L}_{u_k}$ is associated with distinct eigenvalues $u_k$ of $\mathcal{U}$ where $k \in \{0, 1,\dotsc\}$. We define the \textit{symmetry sector} $L_{u_k}$ as the subspace spanned by the eigenvectors of $\mathcal{U}$ associated with the eigenvalue $u_k$. 
We can prove, in close analogy with the Markovian case~\cite{Minganti2018Spectral} that if the steady-state $\dket{\rho_{ss}}$ of~\eqref{def_Leff} is unique, then $\dket{\rho_{ss}}\in L_{u_0=1}$~\cite{Debecker2024PRA}. A spontaneous symmetry breaking (SSB) corresponds to the emergence of a zero eigenvalue in each symmetry sectors $k$ in the limit $N \to \infty$. To be specific, if $\mathcal{L}_{\mathrm{HEOM}}$ is a direct sum of $n+1$ blocks and if its eigenvalues are sorted in each block $k$ as $|\mathrm{Re}[\lambda_0^{(k)}]| < |\mathrm{Re}[\lambda_1^{(k)}]| < \dots$, a SSB is signaled by $\lambda_0^{(k)} \rightarrow \lambda_0^{(0)} = 0 \; \forall k > 0$ for $g\geq g_c, N \rightarrow +\infty$~\cite{Minganti2023D}. This means that the independent hierarchies associated with each block $k$ mix in the limit $N\rightarrow +\infty$ so that steady-states that explicitly break the symmetry emerge.

\begin{figure}
    \centering
\includegraphics[width=0.475\textwidth]{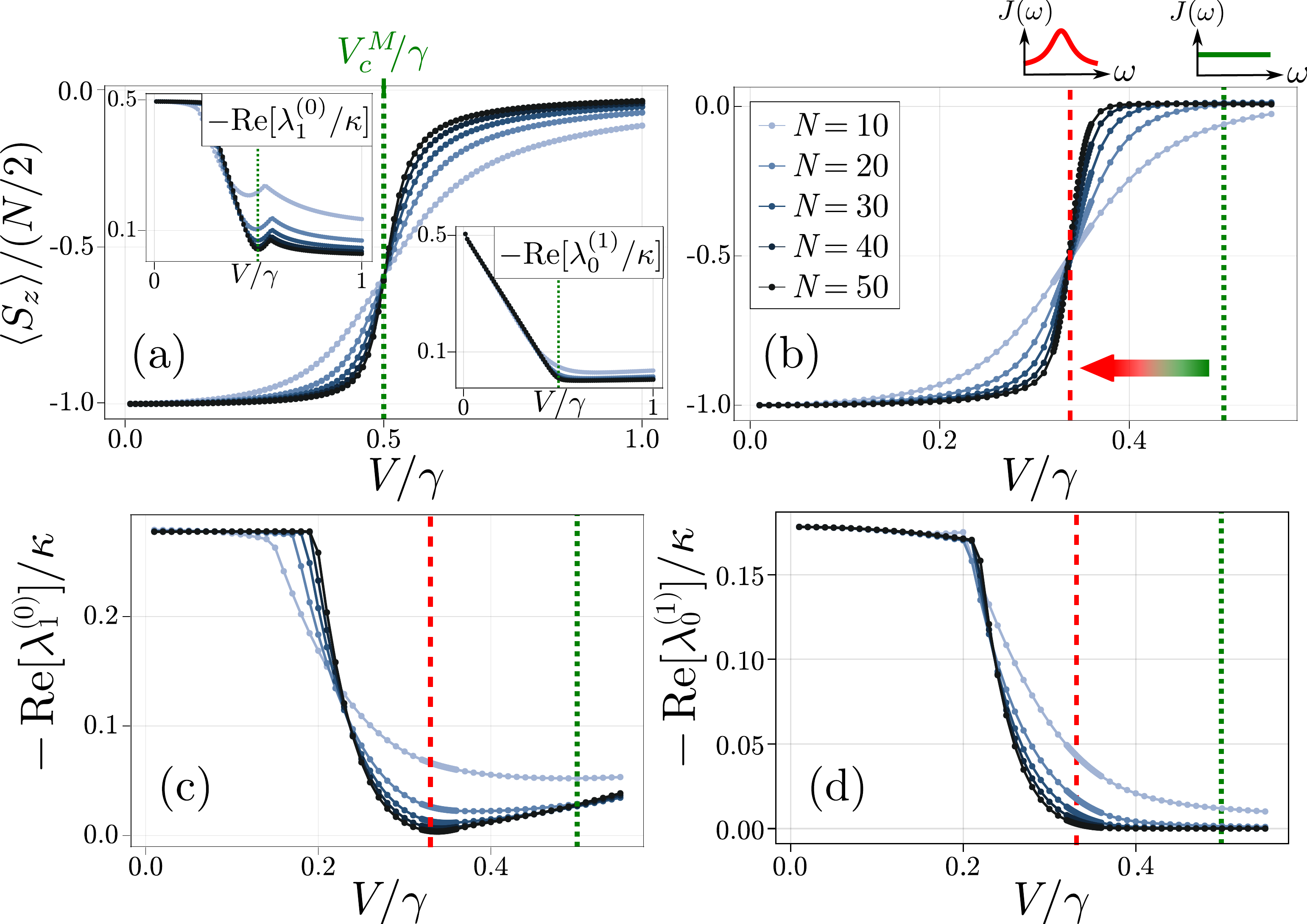}
    \caption{Signatures of the 1$^{\mathrm{st}}$-order DPT for the generalized dissipative LMG model~\eqref{H_LMG} obtained from $\mathcal{L}_{\mathrm{HEOM}}$, showing how environmental spectral structures affect the DPT. (a,b): Steady state magnetization $\braket{S_z}$ as a function of $V/\gamma$ for $\kappa/\omega = 50$ (a) and $\kappa/\omega = 1$ (b). The vertical green and red dashed lines indicate the transitions points for $\kappa/\omega = 50$ and $1$, respectively. (c,d): Liouvillian gap $-\mathrm{Re}[\lambda_1^{(0)}]$ (c) and $-\mathrm{Re}[\lambda_0^{(1)}]$ (d) as a function of $V/\gamma$, indicating respectively the DPT and the SSB associated with the DPT. The insets of (a) show the same quantities for the Markovian case. Truncation orders are $k_\mathrm{max} = 2$ (a) and  $k_\mathrm{max} = 6$ ($N = 10$--$30$), $7$ ($N = 40$), $9$ ($N = 50$) (b-d).}
    \label{1stOrder}
\end{figure}

\emph{1$^{st}$-order DPT.}
We first illustrate our approach for a Lipkin-Meshkov-Glick (LMG) model of the form
\begin{equation}
    H_\mathrm{LMG} = \frac{V}{N}\left(S_x^2 - S_y^2\right) = \frac{V}{2N}\left(S_+^2 + S_-^2\right),
    \label{H_LMG}
\end{equation}
where $S_\alpha = \sum_{j=1}^N \sigma_\alpha^{(j)}/2$ ($\alpha = x, y, z$) are the collective spin operators defined in terms of single-spin Pauli operators $\sigma_\alpha^{(j)}$ and $S_\pm = S_x \pm i S_y$. When the spin system undergoes collective decay as described by Lindblad's master equation
\begin{equation}
   \dot{\rho} = -i[H_\mathrm{LMG}, \rho] + \frac{\gamma}{2N} \mathcal{D}_{S_-}[\rho]
   \label{master_LMG_1}
\end{equation}
where $\mathcal{D}_{o}[\boldsymbol{\cdot}] = 2o\boldsymbol{\cdot} o^\dagger - \{o^\dagger o, \boldsymbol{\cdot}\}$, as would occur if coupled to an unstructured bath with $\alpha(\tau) = (\gamma/N)\delta(\tau)$, the model exhibits a 1$^{st}$-order DPT at the critical point $V_c^{M} = \gamma/2$~\cite{Lee2014D}, separating a steady state phase where $\langle S_z\rangle \to -N/2$ ($V < V_c^{M}$) to a phase where $\langle S_z\rangle \to 0$ ($V > V_c^{M}$) for $N \to \infty$, as can be seen in Fig.~\ref{1stOrder}(a). Here, we generalize the study of this DPT to the non-Markovian regime by considering a finite memory time for the bath with a correlation function of the form $\alpha(\tau) = G\, e^{-\kappa |\tau| -i\omega \tau}$, as if the damping of the collective spin was originating from the coupling of the system to a \textit{structured} bath via an interaction Hamiltonian $H_{\mathrm{int}} =  \sqrt{G}\, (S_-a^\dagger + S_+ a)$ with $G= \gamma \kappa /(2N)$ and $a$ the annihilation operator of a damped pseudo-mode of Hamiltonian $H_E = \omega a^\dagger a$. 
This model allows us to study non-Markovian effects on the DPT and compare them to the Markovian case by tuning the ``loss'' rate $\kappa$ of the pseudo mode. 
Indeed, the collective spin and the pseudo mode form an extended Markovian system governed by the master equation
\begin{equation}
    \dot \rho_\mathrm{tot} = -i[H, \rho_\mathrm{tot}] + \kappa \, \mathcal{D}_a[\rho_\mathrm{tot}]
   \label{master_tot}
\end{equation}
with $H = H_\mathrm{LMG} + H_E + H_\mathrm{int}$. Adiabatic elimination of the pseudo-mode's degrees of freedom recovers Eq.~(\ref{master_LMG_1}) in the limit $\kappa \rightarrow \infty$ (see SM), as expected since $\alpha(\tau) \to (\gamma/N) \delta(\tau)$ for $\kappa \to \infty$. 
When $\kappa$ is finite, memory effects arise and affect the DPT as described below. Note that Eq.~(\ref{master_tot}) has a $\mathbb{Z}_2$ symmetry represented by $\mathcal{U}_2 = U_2 \bcdot U_2^\dagger$ with $U_2 = e^{i\pi (S_z + a^\dagger a)}$. $\mathcal{U}_2$ has two distinct eigenvalues $u_k = e^{ik\pi}=\pm 1$ with $k=0, 1$, so there are two symmetry sectors, with $L_{k = 0}$ containing $\rho_{ss}$. 

The impact of memory effects on the DPT based on $\mathcal{L}_\mathrm{HEOM}$ for the spin system can be seen in Fig.~\ref{1stOrder}. First, we see in panel (b) that the steady state spin magnetization $\langle S_z\rangle$ exhibits a sharp transition at a critical point smaller than in the Markovian case shown in Fig.~\ref{1stOrder}(a). This demonstrates that deviations from a flat spectral density can reshape phase boundaries. A mean-field analysis of~(\ref{master_tot}) shows that the shift in the critical point increases as $\kappa$ decreases (see SM for all details). Physically, this can be understood as follows: the smaller $\kappa$, the greater the probability that excitations escaping from the system will be reabsorbed at later times. The degree of openness of the system therefore decreases as $\kappa$ decreases, which leads to a stabilisation of the phase dominated by the Hamiltonian~(\ref{H_LMG}) for small $V$. For $\kappa \rightarrow 0$ (i.e.,\ for a closed system), the phase transition disappears because the Hamiltonian dynamics no longer competes with dissipative dynamics. In the opposite limit $\kappa \rightarrow \infty$, we recover the Markovian case. The HEOM Liouvillian spectrum correctly captures all DPT signatures. Indeed, it captures the emergence of both the level-touching at the critical point in the symmetry sector $k = 0$, i.e., $-\mathrm{Re}[\lambda_1^{(0)}] \rightarrow 0$ as $N \to \infty$ and the SSB associated to the DPT, i.e., $-\mathrm{Re}[\lambda_0^{(1)}] \rightarrow 0$ for $V$ above the critical point as $N \to \infty$, as can be seen in panels (c) and (d).

Note that this DPT cannot be studied via an approximate reduced description of the spin dynamics obtained after adiabatic elimination of the pseudo-mode, as the related mean-field approach predicts qualitatively different steady states (see SM). In general, reduced descriptions cannot account for all the features of a DPT and can even fail to capture DPTs, as elaborated on further below and strongly motivates again the use of our framework.

\emph{$2^{\mathrm{nd}}$-order DPT}. The second model we consider, also experimentally relevant  for cavity QED~\cite{Morrison2008C}, is of the form~(\ref{GeneralHamiltonian}) with 
$H_S = H_\mathrm{LMG} + h S_z$ and $L = S_x$. For an unstructured bath with $\alpha(\tau) = \gamma \delta(\tau)$, the system dynamics is governed by the master equation
\begin{equation}
   \dot{\rho} = -i[H_\mathrm{LMG} + h S_z, \rho] + \frac{\gamma}{2N} \mathcal{D}_{S_x}[\rho]
   \label{master_LMG}
\end{equation}
whose unique steady-state is the maximally mixed state $\rho_\mathrm{ss} \propto \mathds{1}_{N+1}$, preventing the emergence of any DPT.
However, if we add again a realistic finite memory time for the bath by considering $\alpha(\tau) = (\gamma \kappa/2N) e^{-i\omega \tau- \kappa |\tau|}$, we unveil the existence of two consecutive 2$^{\mathrm{nd}}$-order DPTs separating three different phases upon varying the squeezing strength $V$. This can be seen in Fig.~3, where we show that our approach captures all the features of the DPTs in agreement with mean-field predictions detailed in the SM. Panels (a-c) show the steady states expectations $\braket{S_z}$ and $\braket{S_y^2}$ as a function of $V$, which distinguish the phases [labeled as (I), (II) and (III)], as we have $\braket{S_y^2}$ = 0 in phases (I) and (II) and $\braket{S_z} = -N/2$ in phase (II) only. In addition, as there is a  $\mathbb{Z}_2$ symmetry represented by $\mathcal{U}_2$ in our model akin to the symmetry that is broken in the DPT of the Dicke model~\cite{Damanet2019}, we expect a SSB manifesting as $\lambda_0^{(1)} \rightarrow 0$ as $N\rightarrow +\infty$ in phases (I) and (III), accompanied by an exponential closure of the gap~\cite{Palacino2020}. This behavior is illustrated in panels (b) and (d). Also, note that as the critical points are at $V = -h + \gamma \kappa \omega/[2(\kappa^2 + \omega^2)]$ and $V = h$, taking the limit $\kappa \to \infty$ does not recover the prediction of the Lindblad scenario, i.e., no DPT. In other words, we have $    \mathrm{lim}_{\kappa \to \infty}\mathrm{lim}_{N \to \infty} \neq  \mathrm{lim}_{N \to \infty}\mathrm{lim}_{\kappa \to \infty}$. Physically, taking $\kappa \to \infty$ amounts to consider equal absorption and emission rates for the system, pushing it inevitably to the infinite temperature state. Considering a finite $\kappa$ restores a memory for the bath, i.e., a system-frequency-dependent response, thereby providing the necessary competition between Hamiltonian and dissipative dynamics for the emergence of DPTs.

\begin{figure}
    \centering
\includegraphics[width=0.475\textwidth]{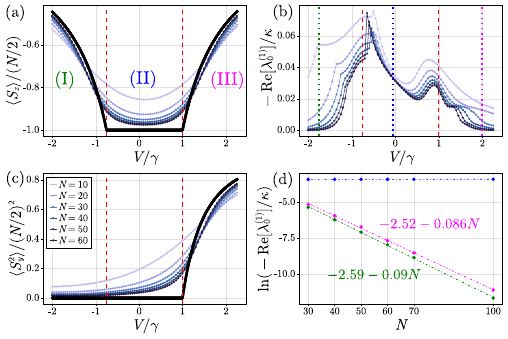}
    \caption{Signatures of the 2$^{\mathrm{nd}}$-order DPT of the second model obtained from $\mathcal{L}_{\mathrm{HEOM}}$, showing the emergence of three phases [(I), (II), and (III)] as $V/\gamma$ is varied. (a) and (c): Steady-state values of $\braket{S_z}$ and $\braket{S_y^2}$ as a function of $V/\gamma$ for different $N$, allowing for distinguishing the phases. The solid black lines correspond to mean-field predictions and the vertical dashed red lines indicate the critical points. (b): Real part of $\lambda_0^{(1)}$(\textit{i.e.,} the gap) as a function of $V/\gamma$, signalling the SSBs associated with the DPT. Three vertical dotted lines indicate the values of $V/\gamma$ taken for the finite-size scaling of the gap shown in panel (d), i.e., $V/\gamma = -1.75$ (green), $-0.05$ (blue), $2$ (pink), revealing an exponential closure of the gap in phases (I) and (III), as shown by the straight line fits. Parameters: $\omega = \kappa = 2 h = 2 \gamma$. Truncation orders: $k_\mathrm{max} = 6$ ($N = 10$-$30$), $7$ ($N = 40$-$60$), $9$ ($N = 70, 100)$.}
    \label{2ndOrder}
\end{figure}

\emph{Conclusion.} 
We applied the spectral theory of non-Markovian dissipative phase transitions developed in the companion paper~\cite{Debecker2024PRA} to demonstrate non-Markovian reservoir engineering of a 1$^{\mathrm{st}}$-order DPT with a discrete SSB and how DPTs can be genuinely triggered by non-Markovian effects.
Our work highlights the importance of exploring out-of-equilibrium matter phases beyond the idealized Markovian limit, featuring non-Markovianity as a resource for triggering or controlling them. This is so far uncharted territory as most works dealing with dissipative many-body dynamics is generally constrained to Lindblad dissipation, which potentially hinders the evidences of DPTs~\cite{Damanet2019, Palacino2020}. 

There are many perspective of our work, such as studies of initial system-bath correlations~\cite{Ikeda2020Generalization} or connections in the non-Markovian regime between DPTs and symmetry breaking~\cite{Huber2020Nonequilibrium, Minganti2021Continuous}, geometric phase curvature~\cite{Carollo2020Geometry, Carollo2018Uhlman}, or dynamical~\cite{Kyaw2020Dynamical, Dolgitzer2021Dynamical} or measurement-induced~\cite{Buchold2021Effective, Muller2022Measurement} phase transitions, or dissipation engineering of long-range order~\cite{Tindall2019}, also for systems with non-Lorentzian environments that exhibit criticality~\cite{Nagy2016Critical, Lundgren2020Nature, Ikeda2020Generalization}.

\begin{acknowledgments}
We thank Jonathan Keeling, Peter Kirton, Valentin Link and Lukas Pausch for
helpful comments on a previous version of the manuscript. Computational resources were provided by the Consortium des Equipements de Calcul Intensif (CECI), funded by the Fonds de la Recherche Scientifique de Belgique (F.R.S.-FNRS) under Grant No. 2.5020.11. 
\end{acknowledgments}

\bibliographystyle{apsrev4-2}
\bibliography{bib}

\begin{thebibliography}{70}
\bibitem [{foo()}]{footnote_formalism}%
  \BibitemOpen
  \href@noop {} {}\bibinfo {note} {It should be noted that the formalism
  presented here is easily generalizable to multiple bosonic or fermionic baths
  at finite temperature.}\BibitemShut {Stop}%
\end{thebibliography}

\end{document}


\title{Supplemental Material: Controlling Matter Phases Beyond Markov}
\author{Baptiste Debecker, John Martin, François Damanet}
\affiliation{Institut de Physique Nucléaire, Atomique et de Spectroscopie, CESAM, University of Liège, B-4000 Liège, Belgium}

\maketitle

\setcounter{equation}{0}
\setcounter{figure}{0}
\setcounter{table}{0}
\setcounter{page}{1}
\makeatletter
\renewcommand{\theequation}{S\arabic{equation}}
\renewcommand{\thefigure}{S\arabic{figure}}
\renewcommand{\bibnumfmt}[1]{[S#1]}
\renewcommand{\citenumfont}[1]{S#1}

This Supplemental Material provides analytical and numerical details on the results presented in the main text of ``Controlling Matter Phases Beyond Markov''. In Sec.~I we provide details on the first example investigated in the main text: the $1^{\mathrm{st}}$-order DPT in the Lipkin-Meshkov-Glick model. We first discuss different descriptions of the system according to different approximations, such as a spin-only description after adiabatic elimination of the pseudomode. Then, we present the mean-field analysis of the spin and of the full (exact) model. We finish the section with the study of the impact of the reservoir parameters on the position of the critical point. Finally, in Sec.~II, we present the mean-field analysis of the second example investigated in the main text: the $2^{\mathrm{nd}}$-order DPTs in the extended LMG model.


\section{First model - Shifting phase boundaries via non-Markovian effects}

In this section, we present details on the first model we consider in the main text, i.e., the generalized dissipative Lipkin-Meshkov-Glick model, which generalizes the study made in Ref.~\cite{Lee2014D} to the non-Markovian regime. The master equation for the collective spin and pseudomode density matrix reads [Eq.~(8) in the main text]
\begin{equation}
\label{ME1DPTv2}
\begin{aligned}
    &\dot\rho_{\mathrm{tot}} = - i \left[H,\rho_\mathrm{tot}\right] + \kappa\left(2 a \rho_\mathrm{tot} a^\dagger - \{ a^\dagger a, \rho_\mathrm{tot}\} \right) \\
&\mathrm{with} \quad \quad 
    H = H_{\mathrm{LMG}} + \omega a^\dagger a + \sqrt{\frac{\gamma \kappa}{2N}} \left( S_- a^\dagger + a S_+\right). \\
     \end{aligned}
\end{equation}
In the following, we first show how adiabatic elimination of the cavity mode recovers the original model [Eq.(7) in the main text] in the ``bad cavity'' limit before performing a mean-field analysis of our generalized model.

\subsection{Adiabatic elimination of the cavity mode in the bad cavity limit}

Let us perform a standard derivation of a master equation for the collective spin only, first dividing $H = H_0 + H_1$ and then working in the interaction picture with respect to $H_0 = H_{\mathrm{LMG}} + \omega a^\dagger a$. In this interaction picture, the interaction Hamiltonian $H_1$ takes the form:
\begin{equation}
    H_{1}(t)
    = \sqrt{\frac{\gamma \kappa}{2 N}} (a(t) S_+(t) + a^\dagger(t) S_-(t)),
\end{equation}
where $S_{\pm}(t) = e^{i H_{\mathrm{LMG}} t} S_{\pm} e^{-i H_{\mathrm{LMG}} t}$. The master equation for the collective spin density operator $\rho = \text{Tr}_{E}(\rho_\mathrm{tot})$ reads in the Markov approximation~\cite{Bre06}:
\begin{equation}
    \dot{\rho} = - \int_0^t dt^\prime \text{Tr}_{E}\left(
    \left[ H_{1}(t), \left[ H_{1}(t^\prime), \rho_\mathrm{tot}(t) \right] \right]
    \right).
\end{equation}
Considering the Born approximation $\rho_\mathrm{tot}(t) \approx \rho(t) \otimes \rho_E$ with $\rho_E$ the vacuum state for the pseudomode and expanding the double commutator
yields
\begin{equation}\label{MEder1}
    \dot{\rho} = - \int_0^t dt' \Big(\alpha(t-t') S_+(t) S_-(t') \rho(t) - \alpha^*(t-t') S_-(t) \rho(t) S_+(t') - \alpha(t-t') S_-(t') \rho(t) S_+(t) + \alpha^*(t-t') \rho(t) S_+(t') S_-(t) \Big),
\end{equation}
which made appear the bath correlation function 
\begin{equation}
    \alpha(t-t') = \left(\frac{\gamma \kappa}{2 N}\right) \text{Tr}_E\left( a(t) a^\dagger(t') \rho_{E} \right)
    = \frac{\gamma \kappa}{2 N} e^{-i \omega (t-t') - \kappa |t-t'|}.
\end{equation}

\paragraph{``Bad cavity'' limit $\kappa \to \infty$.} In the ``bad cavity'' limit $\kappa \to \infty$, which corresponds to the case of a flat spectral density of the bath, we have $\alpha(t-t') \to (\gamma/N) \delta(t-t')$, which makes it possible to perform the integration straightforwardly and obtain the following master equation in the Schr\"odinger picture
\begin{equation} \label{eq:redfield}
\dot{\rho} = - i \left[ H_{\mathrm{LMG}}, \rho\right] +  \frac{\gamma}{2N}\left( 2S_- \rho S_+ - 
  \left\{ S_+ S_-,\rho \right\}\right),
\end{equation}
which is exactly Eq.~(7) in the main text.

\paragraph{Finite $\kappa$ and large detuning limit.} If we know consider the case of finite $\kappa$, the integrand in Eq.~(\ref{MEder1}) is now non-zero over a finite range of time $t'$, so that one has in principle to know the explicit time-dependence of the system operators $S\pm(t') = e^{i H_{\mathrm{LMG}} t} S_{\pm} e^{-i H_{\mathrm{LMG}} t}$ - and thus the $H_\mathrm{LMG}$ Hamiltonian spectrum -- to perform the integral. Such knowledge is however not needed in the limit where the pseudomode frequency $\omega$ dominates all system transition frequencies as shown below. Suppose indeed a spectral decomposition of $H_\mathrm{LMG}$ of the form $H_\mathrm{LMG} |n\rangle = \omega_n |n\rangle$ with eigenvalues and eigenvectors $\omega_n$ and $|n\rangle$. The time-dependence of the system operators $S_\pm(t')$ in the $H_\mathrm{LMG}$ Hamiltonian basis reads
\begin{equation}\label{timedependenceSpm}
    S_\pm(t) = \sum_{n,n'} e^{-i(\omega_n - \omega_{n'})t} \langle n | S_\pm | n'\rangle |n \rangle \langle n'|.
\end{equation}
Inserting Eq.~(\ref{timedependenceSpm}) in Eq.~(\ref{MEder1}), making the variable substitution $t' \to t - \tau$ and pushing the limit of the integral to infinity yields after integration
\begin{equation}
\begin{aligned}
      \dot{\rho} &= - \frac{\gamma \kappa}{2 N} \sum_{n,n'} \Big(\frac{1}{\kappa + i (\omega - (\omega_n - \omega_{n'}))} S_+(t) \left( e^{-i (\omega_n - \omega_{n'})t} \langle n | S_- | n' \rangle | n \rangle \langle n' |  \right)\rho(t)  \Big. \\
       & \quad\quad\quad\quad\quad - \frac{1}{\kappa - i (\omega + (\omega_n - \omega_{n'}))} S_-(t) \rho(t) \left( e^{-i (\omega_n - \omega_{n'})t} \langle n | S_+ | n' \rangle | n \rangle \langle n' |\right) \\
       & \quad\quad\quad\quad\quad - \frac{1}{\kappa + i (\omega - (\omega_n - \omega_{n'}))}  \left( e^{-i (\omega_n - \omega_n')t} \langle n | S_- | n' \rangle | n \rangle \langle n' |\right)  \rho(t) S_+(t) \\
       & \quad\quad\quad\quad\quad \Big. + \frac{1}{\kappa - i (\omega + (\omega_n - \omega_{n'}))}   \rho(t) \left( e^{-i (\omega_n - \omega_{n'})t} \langle n | S_+ | n' \rangle | n \rangle \langle n' |  \right) S_-(t) \Big).
\end{aligned}
\end{equation}
If now we have $\omega \gg (\omega_n - \omega_{n'}) \;\forall n,n'$, then $1/(\kappa \pm i(\omega \pm (\omega_n - \omega_{n'})) \approx 1/(\kappa \pm i\omega)$ and the equation becomes
\begin{equation}
\begin{aligned}
      \dot{\rho} &= - \frac{\gamma \kappa}{2 N} \Big(\frac{1}{\kappa + i \omega} S_+(t) S_-(t) \rho(t) - \frac{1}{\kappa - i \omega} S_-(t) \rho(t) S_+(t) - \frac{1}{\kappa + i \omega}  S_-(t) \rho(t) S_+(t)  + \frac{1}{\kappa - i \omega}   \rho(t) S_+(t) S_-(t) \Big).
\end{aligned}
\end{equation}
Coming back to the Schrödinger picture yields
\begin{equation} \label{eq:largedetuning}
\dot{\rho} = - i \left[ H_{\mathrm{LMG}} -  q_2\frac{\gamma}{2N} S_+ S_-, \rho\right] +  q_1 \frac{\gamma}{2N}\left( 2S_- \rho S_+ - 
  \left\{ S_+ S_-,\rho \right\}\right),
\end{equation}
where we introduced the factors 
\begin{equation}\label{q1q2}
    q_1 =  \frac{\kappa^2}{\kappa^2 + \omega^2}, \qquad q_2 = \frac{\kappa \omega}{\kappa^2 + \omega^2}.
\end{equation}
Hence, we see that in the large detuning limit $\omega/(\omega_n - \omega_{n'}) \to \infty \;\forall n,n'$, the structure of the bath spectral density has the effect of reducing the impact of the dissipative part of the master equation for the spin by a factor $q_1 = \kappa^2/(\kappa^2 + \omega^2)$ as well as adding a energy shift proportional to $q_2 \gamma/2N$. 

\paragraph{Redfield master equation.} As explained above, going beyond the approximations used in the previous section require to take into account the spectrum of the system to evaluate the dissipative part of the master equation. However, the Lipkin-Meshkov-Glick Hamiltonian has complicated eigenvalues and eigenvectors~\cite{Ribeiro2008Exact,Debergh2001} and it is not easy to write down simple formula for Eq.~(\ref{timedependenceSpm}), which hampers the writing of a simple expression for the master equation and stongly motivates the use of our systematic HEOM approach.

\subsection{Mean-field analysis}

We first start by writing the Heisenberg equations of motion for $a$, $S_x$, $S_y$ and $S_z$ from Eq.~(\ref{ME1DPTv2})
\begin{equation}
\begin{aligned}
    &\dot a = -(\kappa + i \omega) a - i \sqrt{\frac{\gamma \kappa}{2 N}} S_-, \\
    &\dot S_x = - \frac{V}{N} \{ S_y, S_z \} + i \sqrt{\frac{\gamma \kappa}{2N}} S_z (a - a^\dagger),\\
    &\dot S_y = - \frac{V}{N} \{ S_x, S_z \} -  \sqrt{\frac{\gamma \kappa}{2N}} S_z (a + a^\dagger),\\
    &\dot S_z = 2 \frac{V}{N} \{ S_x, S_y \} + i \sqrt{\frac{\gamma \kappa}{2 N}} \left( a^\dagger S_- - a S_+\right).
    \end{aligned}
\end{equation}
Using $\langle A B  \rangle = \frac{1}{2}\left\langle \{A,B \} + [A,B]\right\rangle \approx \langle A \rangle \langle B \rangle + \frac{1}{2} \langle [A,B] \rangle$, we get the mean-field semiclassical equations of motion
\begin{align}\label{EOMFull1}
    & \langle \dot a \rangle = -(\kappa + i \omega) \langle a \rangle - i \sqrt{\frac{\gamma \kappa}{2 N}} \langle S_- \rangle, \\ \label{EOMFull2}
    &\langle \dot S_x \rangle = - 2\frac{V}{N} \langle S_y \rangle \langle  S_z \rangle + i \sqrt{\frac{\gamma \kappa}{2N}} \langle S_z\rangle  (\langle a \rangle - \langle a^\dagger \rangle),\\\label{EOMFull3}
    &\langle \dot S_y \rangle = - 2\frac{V}{N} \langle S_x \rangle \langle  S_z \rangle - \sqrt{\frac{\gamma \kappa}{2N}} \langle S_z\rangle  (\langle a \rangle+ \langle a^\dagger \rangle),\\\label{EOMFull4}
    &\langle \dot S_z \rangle = 4\frac{V}{N} \langle S_x \rangle \langle  S_y \rangle + i \sqrt{\frac{\gamma \kappa}{2N}} (\langle a^\dagger \rangle \langle S_-\rangle  - \langle a \rangle \langle S_+\rangle ),
\end{align}
which are exact in the thermodynamic limit $N \to \infty$. Note that the quantity $\langle S_x \rangle^2+\langle S_y \rangle^2+\langle S_z \rangle^2$ is a constant of motion, hence the dynamics of the spin is constrained to the surface of a Bloch sphere. In the following, we are interested in finding the fixed points of these equations and studying their stability as a function of the parameters of the model, in order to deduce the different possible steady states and build a phase diagram as in~\cite{Lee2014D}. 

The different fixed points labelled as $\{(\langle a \rangle, \langle S_x \rangle,\langle S_y \rangle,\langle S_z \rangle)_* \}$ of Eqs.~(\ref{EOMFull1})-(\ref{EOMFull4}) can be obtained by setting the left-hand-side of all the equations to zero. In particular, from the first equation, setting $\langle \dot{a} \rangle = 0$ gives
\begin{equation}\label{adiabatic}
    \langle a \rangle_*  = -i \frac{\sqrt{\frac{\gamma \kappa}{2 N}} \langle S_- \rangle_*}{\kappa + i \omega}
\end{equation}
Hence, the pseudomode degree of freedom is slaved to the collective spin ones, which simplifies the problem as we just need to find first the fixed points of a closed set of equations for the collective spin variables $\langle S_x \rangle_*,\langle S_y \rangle_*$ and $\langle S_z \rangle_*$ only, i.e.,
\begin{align}\label{EOMRedfield1}
    &0 = - 2\frac{V}{N} \langle S_y \rangle_* \langle  S_z \rangle_* + \frac{\gamma}{N} \langle S_z\rangle_* \left(q_1 \langle S_x \rangle_* - q_2 \langle S_y \rangle_*\right),  \\\label{EOMRedfield2}
    &0 = - 2\frac{V}{N} \langle S_x \rangle_* \langle  S_z \rangle_*  + \frac{\gamma}{N} \langle S_z \rangle_* \left(q_1 \langle S_y \rangle_* + q_2 \langle S_x \rangle_*\right), \\\label{EOMRedfield3}
    &0 = 4\frac{V}{N} \langle S_x \rangle_* \langle  S_y \rangle_* - \frac{\gamma}{N} q_1 \left(\langle S_x \rangle^{2}_* + \langle S_y \rangle^{2}_*\right).
    \end{align}
It is worth noting that if one is interested in deriving a spin-only description of the model by performing an adiabatic elimination of the pseudomode (as done in the case of the Dicke model in~\cite{Damanet2019}), it would also require to setting $\dot{a} = 0$ in Eqs.~(\ref{EOMFull1})-(\ref{EOMFull4}), which would yield Eqs.~(\ref{EOMRedfield1})-(\ref{EOMRedfield3}) but with the derivative of the collective spin operators on the left-hand-sides of each equation. Importantly, this means the fixed points of the spin system after adiabatic elimination of the pseudomode are the same as the full model. However, as we will see below, their stability differs depending on whether or not we account for the pseudomode degree of freedom. Note also that taking the limit $\kappa/\omega \to \infty$ in Eqs.~(\ref{EOMRedfield1})-(\ref{EOMRedfield3}) yields $q_1 \to 1$ and $q_2 \to 0$, so that we recover the semiclassical equations of the Lindblad model~\cite{Lee2014D}. 

Equations~(\ref{EOMRedfield1})-(\ref{EOMRedfield3}) together with the normalization condition $\langle S_x\rangle^{2}_* +\langle S_y\rangle^{2}_* + \langle S_z \rangle^{2}_* = (N/2)^2$ admit (as for the Markovian case) two classes of fixed points:
\begin{equation}\label{FixedPoints1}
\begin{aligned}
    \left(\langle S_x \rangle, \langle S_y\rangle, \langle S_z\rangle  \right)_* &= \frac{N}{2}\left(0,0,\pm 1\right)
    \end{aligned}
\end{equation}
and
\begin{align}\label{FixedPoints2a}
   \left(\langle S_x \rangle, \langle S_y\rangle, \langle S_z\rangle  \right)_* &= \frac{N}{2} \left( \frac{\sqrt{1 \pm \sqrt{1-\frac{q_1^2 \gamma^2}{4 V^2}}}}{\sqrt{2}}, 
    \frac{q_1 \gamma}{4V} \frac{\sqrt{2}}{\sqrt{1 \pm \sqrt{1-\frac{q_1^2 \gamma^2}{4 V^2}}}}
    ,0\right), \\\label{FixedPoints2b}
    \phantom{\mathrm{or}} \quad   \left(\langle S_x \rangle, \langle S_y\rangle, \langle S_z\rangle  \right)_* &= -\frac{N}{2} \left(\frac{\sqrt{1 \pm \sqrt{1-\frac{q_1^2 \gamma^2}{4 V^2}}}}{\sqrt{2}}, 
     \frac{q_1\gamma}{4V} \frac{\sqrt{2}}{\sqrt{1 \pm \sqrt{1-\frac{q_1^2 \gamma^2}{4 V^2}}}}
    ,0\right).
\end{align}
According to Eq.~(\ref{adiabatic}), the first class of fixed points corresponds to the spin at the north and south poles of the Bloch sphere with an empty pseudomode, while the second class of fixed points corresponds to the spin at four different locations at the equator and a scaling $\langle a \rangle_* \propto \sqrt{N}$ for the pseudomode. In the following, we analyse the stability of these fixed points to determine the steady states of the model.
As it turns out that it depends on whether the analysis is performed on the spin degrees of freedom $( S_x, S_y, S_z)$ only (spin model), as obtained from adiabatic elimination of the pseudomode, or on all degrees of freedom $(a, S_x, S_y, S_z)$ (full model), we investigate subsequently these two cases. 

\subsubsection{Spin model} 

\paragraph{Linear stability analysis.} Let us first perform a linear stability analysis for the spin model obtained after adiabatic elimination of the pseudomode, i.e.,
\begin{align}\label{EOMRedfield1t}
    &\langle \dot{S}_x \rangle = - 2\frac{V}{N} \langle S_y \rangle \langle  S_z \rangle + \frac{\gamma}{N} \langle S_z\rangle \left(q_1 \langle S_x \rangle - q_2 \langle S_y \rangle\right),  \\\label{EOMRedfield2t}
    &\langle \dot{S}_y \rangle = - 2\frac{V}{N} \langle S_x \rangle \langle  S_z \rangle  + \frac{\gamma}{N} \langle S_z \rangle \left(q_1 \langle S_y \rangle + q_2 \langle S_x \rangle\right), \\\label{EOMRedfield3t}
    &\langle \dot{S}_z \rangle = 4\frac{V}{N} \langle S_x\rangle \langle  S_y \rangle - \frac{\gamma}{N} q_1 \left(\langle S_x \rangle^2 + \langle S_y \rangle^2\right),
    \end{align}
around the first class of fixed points: $\left(\langle S_x \rangle, \langle S_y\rangle, \langle S_z\rangle  \right)_* = \left(0,0,\pm N/2\right)$. Note that the equations above can also be obtained from the master equation~(\ref{eq:largedetuning}) in the large detuning limit. To perform the analysis, we replace $\left(\langle S_x \rangle, \langle S_y\rangle, \langle S_z\rangle  \right)$ by $(x,y,z \pm (N/2))$ in Eqs.~(\ref{EOMRedfield1t}-\ref{EOMRedfield3t}) where $(x,y,z)$ denotes fluctuations around the fixed points which yields
\begin{align}\label{EOMRedfield1BIS}
    &\dot{x} = - 2\frac{V}{N} y \left(z \pm \frac{N}{2}\right) + \frac{\gamma}{N} \left(z \pm \frac{N}{2}\right) \left(q_1 x - q_2 y\right),  \\\label{EOMRedfield2BIS}
    &\dot{y}= - 2\frac{V}{N} x \left(z \pm \frac{N}{2}\right)  + \frac{\gamma}{N} \left(z \pm \frac{N}{2}\right) \left(q_1 y + q_2 x\right), \\\label{EOMRedfield3BIS}
    &\dot{z}=  4\frac{V}{N} x y - \frac{\gamma}{N} q_1 \left(x^2 + y^2\right).
    \end{align}
Linearizing Eqs.~(\ref{EOMRedfield1BIS})-(\ref{EOMRedfield3BIS}) shows that fluctuations along $z$ decouple from those along $x$ and $y$, which reduces the analysis to the following two-dimensional system
\begin{equation}\label{lsSpin}
    \begin{pmatrix}\dot x \\
    \dot y \end{pmatrix} = \pm \frac{\gamma}{2} \begin{pmatrix}
        q_1 & -\frac{2V}{\gamma} - q_2  \\
        -\frac{2V}{\gamma} + q_2 & q_1  \\
    \end{pmatrix} \begin{pmatrix} x \\
    y \end{pmatrix}.
\end{equation}
The eigenvalues of the matrix are $\pm \gamma(q_1-\sqrt{4 (V/\gamma)^2-q_2^2})/2$ and $\pm \gamma(q_1 + \sqrt{4 (V/\gamma)^2-q_2^2})/2$. For the fixed point $\left(\langle S_x \rangle, \langle S_y\rangle, \langle S_z\rangle  \right)_* = \left(0,0,\frac{N}{2}\right)$ [``+'' sign in Eq.~(\ref{lsSpin})], the eigenvalue $\gamma(q_1 +\sqrt{4 (V/\gamma)^2-q_2^2})/2$ is always positive, meaning that the fixed point is always unstable. For the fixed point $\left(\langle S_x \rangle, \langle S_y\rangle, \langle S_z\rangle  \right)_* = \left(0,0,-\frac{N}{2}\right)$ [``--' sign in Eq.~(\ref{lsSpin})], both the eigenvalues $-\gamma(q_1 -\sqrt{4 (V/\gamma)^2-q_2^2})/2$ and $-\gamma(q_1 +\sqrt{4 (V/\gamma)^2-q_2^2})/2$ are negative for 
\begin{equation}\label{Vcp}
    V < \frac{\gamma}{2\sqrt{1 + \omega^2/\kappa^2}} \equiv V_c^{+},
\end{equation}
and $-\gamma(q_1 -\sqrt{4 (V/\gamma)^2-q_2^2})/2$ is positive for $V > V_c^{+}$. Hence, the fixed point $\left(\langle S_x \rangle, \langle S_y\rangle, \langle S_z\rangle  \right)_* = \left(0,0,-\frac{N}{2}\right)$ is stable only if $V < V_c^{+}$ and unstable otherwise. Note again that in the limit $\kappa/\omega \to \infty$, we recover the result of the Lindblad model of~\cite{Lee2014D} where $V_c^{+} \to \gamma/2$.

Let us now perform a linear stability analysis around the second class of fixed points of the form $\left(\langle S_x \rangle, \langle S_y \rangle, \langle S_z \rangle  \right)_* = \left(s_{x*},s_{y*},0\right)$, where $s_{x*}$ and $s_{y*}$ are given by the right-hand side of Eqs.~(\ref{FixedPoints2a}) or~(\ref{FixedPoints2b}). Here, the EOM for the fluctuations, obtained by replacing $\left(\langle S_x \rangle, \langle S_y\rangle, \langle S_z\rangle  \right)$ by $(x + s_{x*},y + s_{y*} ,z)$, yields after linearization
\begin{equation}
\begin{pmatrix}\dot x \\
    \dot y \\
    \dot z\end{pmatrix}  = \frac{1}{N}
\begin{pmatrix}
 0 & 0 & q_1 \gamma s_{x*} - (2V + q_2 \gamma) s_{y*} \\
 0 & 0 & q_1 \gamma s_{y*} - (2V - q_2 \gamma) s_{x*} \\
 4V s_{y*}- 2 q_1 \gamma s_{x*} &  4V s_{x*}- 2 q_1 \gamma s_{y*}  & 0 \\
\end{pmatrix}\begin{pmatrix} x \\
    y \\
    z\end{pmatrix} + \begin{pmatrix} 0 \\
    0 \\
    - \frac{N \gamma q_1}{4}\end{pmatrix}.
\end{equation}
The eigenvalues of the matrix are
\begin{equation}\label{eigenvaluesR}
    0,  \quad    -N \sqrt{(\pm q_2 \sqrt{4 V^2- \gamma^2 q_1^2} + \gamma^2 q_1^2 - 4 V^2)/2}, \quad\mathrm{and}\quad
    N \sqrt{(\pm q_2 \sqrt{4 V^2- \gamma^2 q_1^2} + \gamma^2 q_1^2 - 4 V^2)/2},
\end{equation}
where the $\pm$ sign relate to the $\pm$ sign in Eqs.~(\ref{FixedPoints2a}) and (\ref{FixedPoints2b}). Note that the zero eigenvalue is not physical, as the variables are not independent. To see how this eigenvalue can be eliminated, one can switch to the spherical coordinates $\langle S_x \rangle = r \sin\theta \cos\varphi$, $\langle S_y \rangle = r \sin\theta \sin\varphi, \langle S_z \rangle = r \cos\theta$ in Eqs.~(\ref{EOMRedfield1t}-\ref{EOMRedfield3t}), which yields $\dot{r} = 0$ because of the norm conservation and 
\begin{align}
&\dot{\varphi} = \frac{\cos\theta}{2}\left[ q_2 \gamma - 2 V \cos(2 \varphi)\right], \\
&\dot{\theta} = \frac{\sin\theta}{2}\left[ q_1 \gamma - 2 V \sin(2 \varphi)\right].
\end{align}
Linear stability analysis based on these equations yields the two non-trivial eigenvalues of Eq.~(\ref{eigenvaluesR}). The fixed points are unstable when the real part of at least one of them is positive. This happens for
\begin{equation}
    V < \frac{\gamma}{2\left( 1 + \omega^2/\kappa^2\right)} \equiv V_c^{-}
\end{equation}
for the two fixed points associated with the ``--'' sign in Eqs.~(\ref{FixedPoints2a}) and (\ref{FixedPoints2b}) and for
\begin{equation}
    V <  V_c^{+}
\end{equation}
for the two fixed points associated with the ``$+$'' sign. Above these critical values, the real part of both the eigenvalues are zero, and so one cannot conclude on the stability of the fixed points in this region via this simple linear stability analysis.

While the approach above cannot provide a full understanding of the stability of the system, we already see strong deviations from the Markovian limit $\omega/\kappa \to 0$. Indeed, in this latter, we have $V_c^{-} = V_c^{+} = V_c^{M} = \gamma/2$ so that there is no region of parameters where the two kind of fixed points are not unstable, hampering the possibility for a coexistence of two distinct phases. More specifically, in~\cite{Lee2014D}, it was shown that for $V < V_c^M$ the fixed point $\left(0,0,-N/2\right)$ is the unique steady state while for $V > V_c^M$ the fixed points $\left(s_{x*},s_{y*},0\right)$ are center fixed points and there exists an infinite set of oscillating (initial-state-dependent) steady states corresponding to orbits around these fixed points on the Bloch sphere. While these steady states are persistent spin oscillations, they average to zero along the $z$ direction over time, so that the Markovian scenario well-describes a $1^{\mathrm{st}}$-order transition between a phase with $\langle S_z \rangle \neq 0$ ($V < V_c^M$) and $\langle S_z \rangle = 0$ ($V > V_c^M$). In the case $\omega/\kappa \neq 0$, however, we have $V_c^{-} < V_c^{+}$, so the possibility that the fixed points and thus two distinct phases coexist in the region $V_c^{-} < V < V_c^{+}$ (which increases as $\omega/\kappa$ increases) is not excluded. This is confirmed in the next subsection.

\paragraph{Higher-order stability analysis.} As the equations of motion are cumbersome beyond linearization, we rely on a numerical analysis of the stability of the fixed points, as summarized below. 

\begin{figure}
    \centering
    \includegraphics[width=0.98\textwidth]{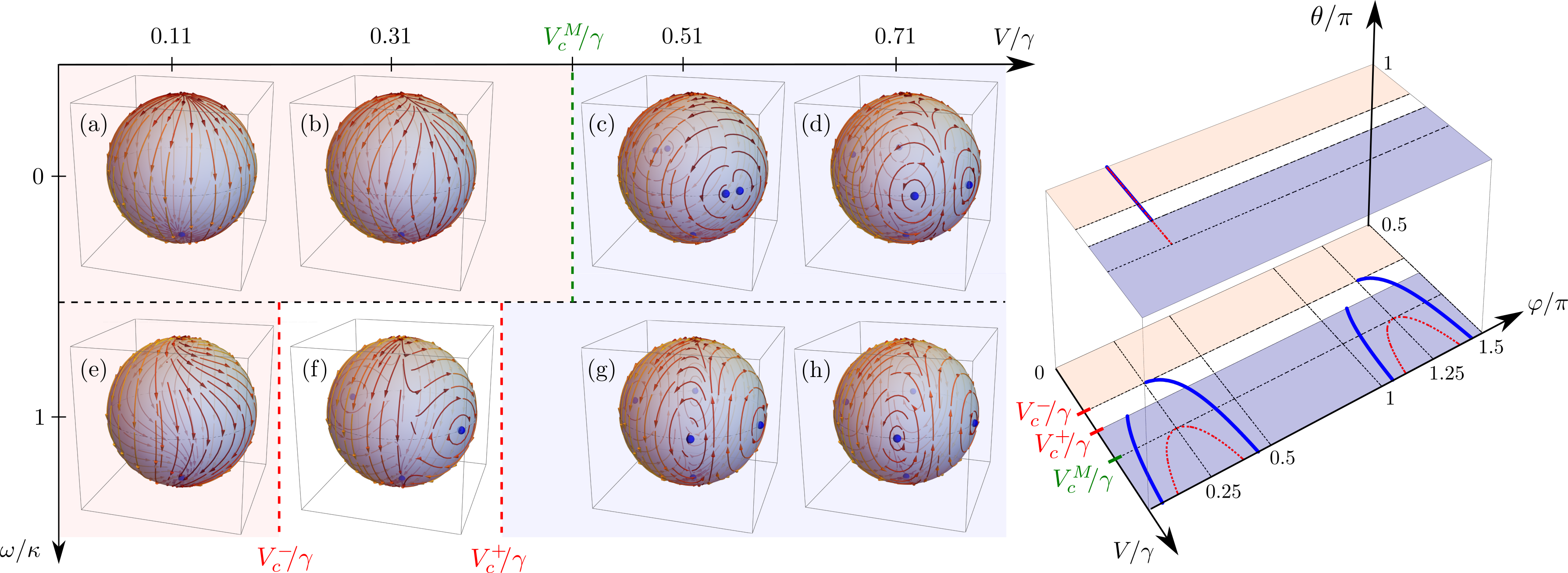}
    \caption{Stream plots obtained from the mean-field equations~(\ref{EOMRedfield1t})-(\ref{EOMRedfield3t}) after adiabatic elimination of the pseudomode showing the trajectories of the collective spin on the Bloch sphere in the thermodynamic limit as a function of the ratio $V/\gamma$ between coherent and dissipative rates (not in scale) and of the degree of spectral density structures $\omega/\kappa$. In the Markovian limit ($\omega/\kappa \to 0$), there is a phase transition at $V = V_c^M = \gamma/2$ between a phase with a unique pure steady state $(0,0,-N/2)$ (blue dot at the south pole) [$V < V^M_c$, (a,b)] and a phase with an infinite set of (initial-state-dependent) pure steady states orbiting around four center fixed points (blue dots at the equator) [$V > V^M_c$, (c,d)]. As $\omega/\kappa$ increases, the stream lines twist around the $z$-axis (e) and a region of parameters ($V_c^{-} < V < V_c^{+}$) emerges where both the steady state $(0,0,-N/2)$ or a steady orbit can be observed depending on the initial condition (f). Overall, the phase space where the coherent dynamics dominates over the dissipative one is enlarged due to non-Markovian effects. Parameters are $V/\gamma = 0.11$ (a,e), $V/\gamma = 0.31$ (b,f), $V/\gamma = 0.51$ (c,g) and $V/\gamma = 0.71$ (d,h). For $\omega/\kappa = 1$, $V_c^{-}/\gamma = 0.25$ and $V_c^{+}/\gamma \approx 0.354$. (i) Positions of the relevant fixed points on the Bloch sphere as a function of spherical coordinates $\varphi$ and $\theta$ and $V/\gamma$ for the Markovian limit $\omega/\kappa = 0$ (dashed red) and for $\omega/\kappa = 1$ (blue).}
    \label{stream}
\end{figure}

For $0 \leq V < V_c^{-}$, the fixed point $\left(0,0,-N/2\right)$ is the unique steady state, as shown in the stream plots in Fig.~\ref{stream} (a,b,e). Note that in this regime all the fixed points of the form $\left(s_{x*},s_{y*},0\right)$ are not physical fixed points as they are complex vectors.

For $V_c^{-} <  V < V_c^{+}$, we observe a bistability phenomenon as the steady state depends on the initial conditions: the fixed point $\left(0,0,-N/2\right)$ is still a possible steady state, but an infinite number of steady states corresponding to stable orbits on the Bloch sphere around the two center fixed points~(\ref{FixedPoints2a}) and (\ref{FixedPoints2b}) with the ``--'' sign become available, as shown in Fig.~\ref{stream}(f). For $V \gtrsim V_c^{-}$, only orbits close to the fixed points are stable. As $V$ increases, more orbits become stable. This is in sharp contrast with the Markovian case where all the points on the sphere belong to stable orbits as soon as $V > V_c^M$ [panel (c,d)]. Here, in the regime $V_c^{-} < V < V_c^{+}$, each point on the Bloch sphere undergoes the phase transition at a different critical point. 

For $V > V_c^{+}$, the fixed point $\left(0,0,-N/2\right)$ is no longer stable and all the four fixed points (\ref{FixedPoints2a}) and (\ref{FixedPoints2b}) become center fixed points, as can be seen in Fig.~\ref{stream}(g,h). Note that since $V_c^{+} < V_c^M$, this means going beyond the limit of a flat spectral density has the effect to renormalize to lower values the critical point by a function of the bath SD. We interpret this phenomenon as a consequence of memory effects: for smaller $\kappa/\omega$, the excitations escaping the system are more likely to be re-absorbed by the system at later times, protecting its coherence and hence stabilizing the phase dominated by the Hamiltonian term to lower ratio $V/\gamma$.

\subsubsection{Full model}

We now come back to the mean-field equations for the full model~(\ref{EOMFull1})-(\ref{EOMFull4}) in order to study how reintroducing the pseudomode degree of freedom affects the stability of the fixed points. Here, the complex variable $\langle a \rangle$ can be decomposed into two real variables. Hence, the linear stability analysis on the full model exhibits four non-trivial eigenvalues to analyze, by contrast with the two of the previous section. As before, the analysis consists in linearizing the equations of motion for the fluctuations around the different fixed points and inspecting the eigenvalues of the matrix of the system. We summarize below the results for each class of fixed points.

\begin{figure}
    \centering
    \includegraphics[width=0.98\textwidth]{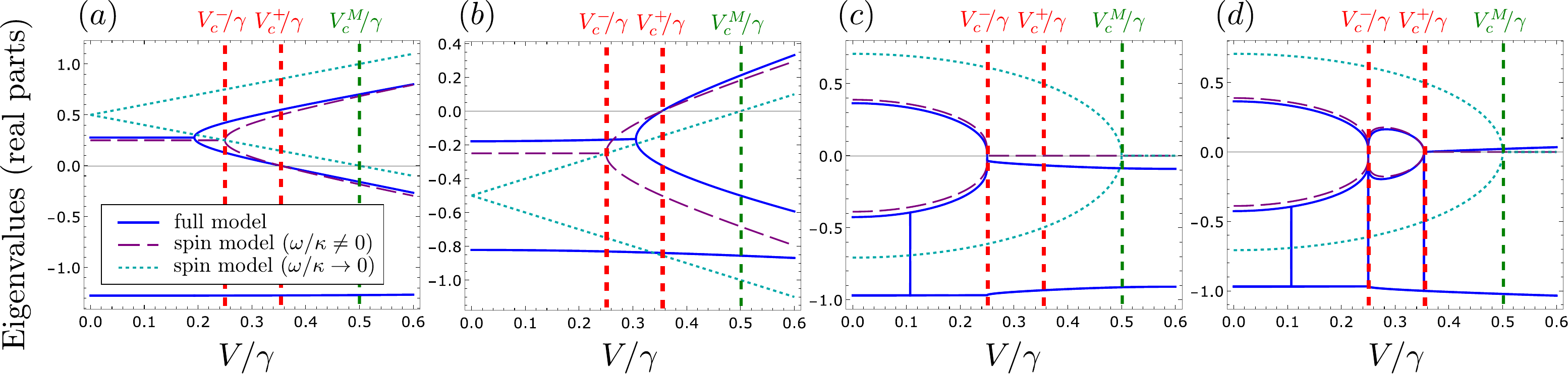}
    \caption{Real parts of the eigenvalues of the linear stability analysis matrix of the mean-field equations for the full model [Eqs~(\ref{EOMFull1})-(\ref{EOMFull4})] (blue solid lines) with $\omega = \kappa = \gamma$ and for the spin model after adiabatic elimination of the pseudomode [Eqs.~(\ref{EOMRedfield1t})-(\ref{EOMRedfield3t})] (dashed purple lines) with $\omega = \kappa = \gamma$ and in the limit $\omega/\kappa \to 0$ (dotted dark cyan lines). The different panels correspond to linear stability analyses around different fixed points: at the north pole of the Bloch sphere [$\left(\langle S_x \rangle, \langle S_y\rangle, \langle S_z\rangle  \right)_* = \left(0,0, -N/2\right)$] (a), at the south pole [$\left(0,0, -N/2\right)$] (b) and at the equators [Eqs.~(\ref{FixedPoints2a})-(\ref{FixedPoints2b}) with the ``--'' sign (c) and the ``+'' sign (d)]. For a given $V/\gamma$, the fixed points are stable if all the corresponding eigenvalues have a negative real parts, while they are unstable if at least one of them is positive. The vertical dashed lines indicate the positions of the special values of $V/\gamma$ as described in the text.}
    \label{fig:eigenvalues}
\end{figure}

For the first class of fixed points at the north and south poles of the Bloch sphere, $\left(\langle S_x \rangle_*, \langle S_y \rangle_*, \langle S_z \rangle_*  \right) = \left(0,0, N/2\right)$ is still unstable for all values of $V$, while $\left(0,0, -N/2\right)$ is still a valid steady state for $V < V_c^+$, as can be seen in panel (a) and (b) respectively of Fig.~\ref{fig:eigenvalues}, which displays the real parts of the eigenvalues of the linear stability analysis matrix in each case.

For the second class of fixed points at the equator of the Bloch sphere, the changes are more drastic. (i) The two fixed points associated with the ``--'' sign in Eqs.~(\ref{FixedPoints2a}) and (\ref{FixedPoints2b}) are still unstable for $V <  V_c^{-}$. However, they are now \textit{stable} for $V > V_c^{-}$, as can be seen in Fig.~\ref{fig:eigenvalues}(c) showing that all real parts of the eigenvalues become negative in this region (blue solid curve), by contrast with the analysis of the reduced spin model (dashed purple and dotted dark cyan). Hence, these two fixed points are now real steady states for $V > V_c^{-}$, and they coexist with the steady state $\left(0,0, -N/2\right)$ for $V_c^{-} < V < V_c^{+}$. There are no more stable orbits around these fixed points. (ii) The two fixed points associated with the ``+'' sign in Eqs.~(\ref{FixedPoints2a}) and (\ref{FixedPoints2b}) are still unstable for $V <  V_c^{+}$, and now become also unstable above $V_c^{+}$, as can be seen in Fig.~\ref{fig:eigenvalues}(d). 

In conclusion, the mean-field analysis of our generalized LMG model~(\ref{ME1DPTv2}) yields qualitatively different phase diagrams depending on the approximations used. 
\begin{itemize}
    \item The original spin-only model of~\cite{Lee2014D} as obtained in the limit $\omega/\kappa \to 0$ exhibits the DPT at $V = V_c^M = \gamma/2$. For $V < \gamma/2$, $\left(0,0, -N/2\right)$ is the only steady state, while for $V > \gamma/2$ they are four center fixed points at the equator of the Bloch sphere around which an infinity of stable orbits are valid steady states, which one is chosen depends on the initial conditions.
    \item When including the spectral structures of the bath in the reduced description of the spin system ($\omega/\kappa \neq 0$), the critical point is split into two: $V_c^- = \gamma/[2(1+\omega^2/\kappa^2)]$ and $V_c^+ = \gamma/(2\sqrt{1+\omega^2/\kappa^2})$ where $V_c^- < V_c^+$, and the four fixed points at the equator have shifted locations. For $V < V_c^-$, $\left(0,0, -N/2\right)$ is the only steady state. For $V_c^- < V < V_c^+$, $\left(0,0, -N/2\right)$ and an ensemble of stable orbits around two out of the four center fixed points can be steady states. For $V > V_c^+$, only stable orbits around the four center fixed points are steady states.
    \item In the exact full model, the fixed points remain the same but not their stability. Notably, stable orbit cannot be seen anymore in any regime. For $V < V_c^-$, $\left(0,0, -N/2\right)$ is the only steady state. For $V_c^- < V < V_c^+$, $\left(0,0, -N/2\right)$ and two fixed points at the equator can be steady states. For $V > V_c^+$, only these two latter can be steady states. 
\end{itemize}
Signatures of this overall picture can be found within our HEOM approach, as it predicts for finite $N$ a transition between $V_c^-$ and $V_c^+$. We elaborate on this in the next section.

\subsection{Impact of non-Markovianity on the critical point}
In the main text, we showed that for the first LMG model we consider one can reshape the phase boundaries by considering a bath correlation function that decays with a finite time $1/\kappa$. In this section, we study more thoroughly the impact of this finite decay time on the DPT. 

\begin{figure}
    \centering
    \includegraphics[width=0.98\textwidth]{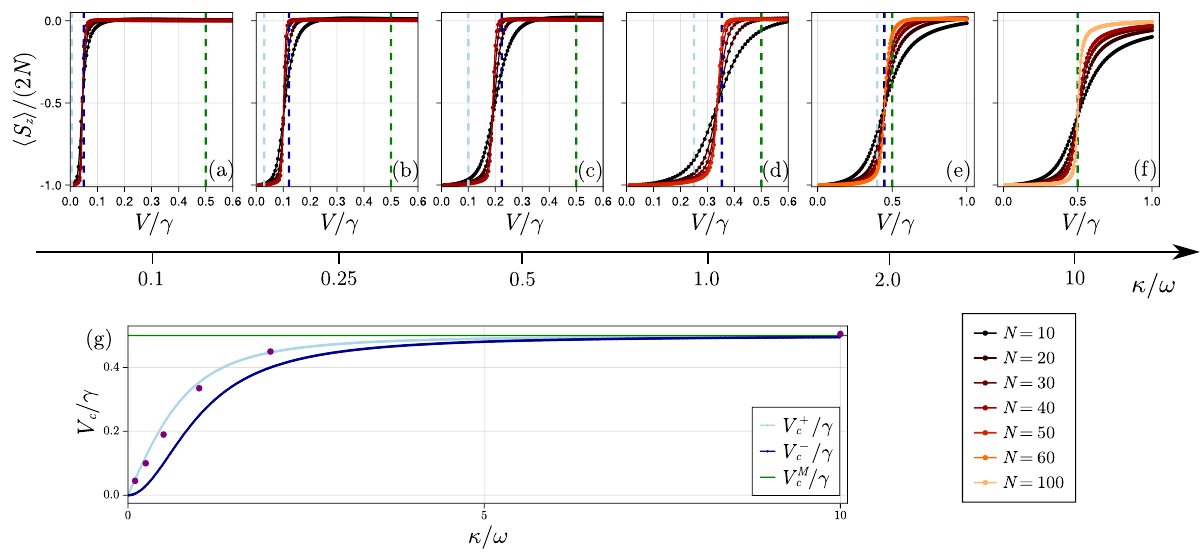}
    \caption{Panels (a)-(f): Magnetization $\braket{S_z}/(2N)$ as a function of $V/\gamma$ for different values of $\kappa/\omega$ [see axis below the plots] and for different values of $N$ [see legend] highlighting the reshaping of the phase boundaries induced by non-Markovianity. Panel (g): Effective critical point $V_c/\gamma$ (dots in purple), \textit{i.e.} the critical point numerically found from $\mathcal{L}_\mathrm{HEOM}$, as a function of $\kappa/\omega$. We see that $V_c/\gamma$ is always between $V_c^+\gamma$ and $V_c^-/\gamma$ and appears to closely follow the scaling of $V_c^+/\gamma$ given in Eq.~\eqref{Vcp}.} 
    \label{ImpactOfkappa}
\end{figure}

We find that the smaller $\kappa/\omega$, the sharper the transition for a given $N$ as illustrated by the panels (a)-(f) of Fig.~\ref{ImpactOfkappa}. This result is not surprising as lowering $\kappa/\omega$ takes us further from the Markovian limit  defined by $\kappa \rightarrow +\infty$, which in turn leads to larger values of $k_\mathrm{max}$ to ensure convergence. However, as $k_\mathrm{max}$ grows, the dimension of the HEOM Liouvillian $\mathcal{L}_\mathrm{HEOM}$ also increases, in close analogy with the thermodynamic limit $N \rightarrow +\infty$ which also increases the dimension of the HEOM Liouvillian.
The panel (g) of Fig.~\ref{ImpactOfkappa} shows that the critical point, numerically extracted from $\mathcal{L}_\mathrm{HEOM}$ is always between $V_c^+$ and $V_c^-$ for all values of $\kappa/\omega$ considered here, as expected. Furthermore, the critical point $V_c/\gamma$ seems to more closely follow the scaling of $V_c^+/\gamma$ given in Eq.~\eqref{Vcp} than that of $V_c^-/\gamma$.

\section{Second model - Triggering DPTs via non-Markovian effects} 

This section is concerned  with the second model considered in the main text, i.e., the modified LMG model coupled to a bath with exponentially decaying correlation function, which can be modelled as a spin system coupled to a damped pseudomode $a$.
\subsection{Mean-field analysis}
In this section, we analyze the mean-field equations associated with the second model considered in the main text. The master equation for total density matrix reads 
\begin{equation}
 \begin{aligned}
   &\dot{\rho}_\mathrm{tot} = -i[H, \rho_\mathrm{tot}] + \kappa\left(2 a \rho_\mathrm{tot} a^\dagger - \{ a^\dagger a, \rho_\mathrm{tot}\} \right)  \\
&\mathrm{with} \quad \quad  H = H_\mathrm{LMG} + h S_z + \omega a^\dagger a + \sqrt{\frac{\gamma \kappa}{2N}}S_x(a+a^\dagger).
  \end{aligned}
   \label{master_LMG_tot_2}
\end{equation}

In close analogy with Section IV. B., it is easy to show that the mean-field equations read 
\begin{align}\label{EOMFull_12}
     \langle \dot a \rangle &= -(\kappa + i \omega) \langle a \rangle - i \sqrt{\frac{\gamma \kappa}{2 N}} \langle S_x \rangle, \\ \label{EOMFull22}
    \langle \dot S_x \rangle &= - 2\frac{V}{N} \langle S_y \rangle \langle  S_z \rangle - h \braket{S_y},\\\label{EOMFull32}
    \langle \dot S_y \rangle &= - 2\frac{V}{N} \langle S_x \rangle \langle  S_z \rangle + h \braket{S_x} - \sqrt{\frac{\gamma \kappa}{2N}} \langle S_z\rangle  (\langle a \rangle+ \langle a^\dagger \rangle),\\\label{EOMFull42}
    \langle \dot S_z \rangle &= 4\frac{V}{N} \langle S_x \rangle \langle  S_y \rangle +  \sqrt{\frac{\gamma \kappa}{2N}} \braket{S_y} (\braket{a} + \braket{a^\dagger}).
\end{align}
They have six fixed points from which four phases labelled as (I), (II), (IIb) and (III) can be inferred:
\begin{align}
&\text{(I): } \left(\langle a \rangle, \langle S_x \rangle, \langle S_y\rangle, \langle S_z\rangle  \right)_* = \frac{N}{2} \left(\mp \sqrt{\frac{\gamma}{2 N \kappa}}\sqrt{1 - \left(\frac{ h}{V - \frac{q_2 \gamma}{2}}\right)^2} (q_2 + i q_1), \pm \sqrt{1 - \left(\frac{ h}{V - \frac{q_2 \gamma}{2}}\right)^2},0, \frac{h}{\left(V-\frac{q_2 \gamma}{2}\right)}\right) \\
&\text{(II): } \left(\langle a \rangle, \langle S_x \rangle, \langle S_y\rangle, \langle S_z\rangle  \right)_* = \frac{N}{2}\left(0,0,0, -1\right) \\
&\text{(IIb): } \left(\langle a \rangle, \langle S_x \rangle, \langle S_y\rangle, \langle S_z\rangle  \right)_* = \frac{N}{2}\left(0,0,0, 1\right) \\
&\text{(III): } \left(\langle a \rangle, \langle S_x \rangle, \langle S_y\rangle, \langle S_z\rangle  \right)_* =  \frac{N}{2}\left(0,0,\pm \sqrt{1 - \left(\frac{h}{V}\right)^2}, -\frac{h}{V}\right).
\label{MF-phaseIII}
\end{align}
If $h > 0$ ($h < 0)$, the phase (IIb) [(II)] is always unstable. As we chose $h > 0$ in the main text, we thus only have the phases (I), (II), and (III) to consider. Phases (I) and (III) both gather two fixed points, corresponding in each case to two broken symmetry states.  Also, we note that their fixed points are unphysical for $| V - q_2\gamma/2| < h $ and $|V| < h$, respectively, which already gives some hints about their stability. A linear stability analysis around the fixed points show that the phases are stable in distinct regions of parameters. For $h, q_2,\gamma > 0$ as considered in the main text,  we have the two critical points $V_1 = \mathrm{min}\left(-h + \frac{q_2 \gamma}{2},\frac{q_2 \gamma}{4} \right)$ and $V_2 = \mathrm{max}\left(h,  \frac{q_2 \gamma}{4} \right)$ with phase (I) stable for $V < V_1$; phase (II) stable for $V_1 < V < V_2$; and phase (III) stable for $V_2 < V$. For $h > q_2 \gamma/4$, the two critical points coalesce and phase (II) does not emerge anymore when varying $V$. In the particular case $\omega = \kappa = 2 \gamma = 2 h$ as considered in the main text, we have $V_1 = -3\gamma/4$ and $V_2 = \gamma$.

\bibliographystyle{apsrev4-2}
\bibliography{bib}